# Long memory in select stock returns using an alternative wavelet log-scale alignment approach

Avishek Bhandari[1] and Bandi Kamaiah[2]

## Abstract

This study investigates the efficiency of some select stock markets. Using an improved wavelet estimator of long range dependence, we show evidence of long memory in the stock returns of some emerging Asian economies. However, developed markets of Europe and the United States did not exhibit long memory thereby confirming the efficiency of developed stock markets. On the other hand, emerging Asian markets are found to be less efficient as long memory is more pronounced in these markets.

## 1.1 Introduction

Long memory processes, also known as long-range dependent process, are ubiquitous in financial and economic time-series. This study seeks to understand the long memory behaviour of global equity returns using novel methods from wavelet analysis, where long-run correlation structure of major global equity returns are analysed within the framework of wavelet log-scale method. The first study of long memory was conducted by Hurst (1951) when analysing the flow of Nile River. Mandelbrot and Van Ness (1968), using the idea of Hurst exponent, employed the idea of long-memory processes in conjunction with fractional Brownian motion and related stochastic processes. However, in the field of time series analysis, Granger and Joyeux (1980) and Hosking (1981) were among the first to integrate long memory processes with time series methods. Majority of research focusing on estimation of long memory relies on the traditional rescaled range (R/S) approaches of Mandelbrot and Wallis (1968) and its modified version developed by Lo (1991). The spectral domain approach proposed by Geweke and Porter-Hudak (1983) to estimate the long memory parameter has been used by many researchers too.

---

[1] Institute of management technology, Hyderabad, India, 501218, Email: bavisek@gmail.com
[2] Professor emeritus, University of Hyderabad, India, 500046, Email: kamaiahbandi@gmail.com

This paper investigates long memory among select global equity markets using estimators from the wavelet domain. Studies investigating long memory in global financial markets based on wavelet based long memory methods are relatively fewer as compared to traditional time and spectral domain estimators of long memory. Furthermore, empirical studies based on log-scale wavelet domain estimator of long-range dependence are practically nonexistent. The dearth of studies concerning wavelet based analysis of long memory of global equity markets necessitates an exploration based on multiscale wavelet domain methods. To the best of our knowledge, this is the first study on long memory of equity returns that employs the wavelet log-scale method.

## 1.2 Literature Review

The presence of long memory in squared daily returns of S&P 500 index is evident in the works of Ding et al. (1993) where significant autocorrelation for lags up to ten years were present. Similarly, Lobato and Savin (1998) also demonstrated the presence of long memory in the squared returns of the S&P 500 dataset spanning three decades. Furthermore, Lobato and Velasco (2000) using a frequency domain tapering procedure in a multi stage semi-parametric method unearthed the presence of long memory in stock returns and volatility of returns. The presence of long memory in the returns of Brazilian equity market is documented in Assaf and Cavalcante (2005). Barkoulas et al. (2000), while investigating the long memory properties of the Athens stock exchange, find evidence of long-range persistence in the returns of the Athens stock market. Moreover, the forecast performance of a long memory incorporated model significantly outdid forecasts generated from a regular random walk model. Similarly, Panas (2001), using a spectral measure of fractality along with the Levy index, found nonlinearities in Greek equity returns and unearthed the existence of long memory, thereby rejecting the weak-form efficiency of the Greek equity market. Henry (2002), using a mixture of semi-parametric and spectral estimators, found evidence of long memory in the returns of South Korean stock market. Moreover, some evidence of weak long memory was unearthed in the markets of Germany and Taiwan. The empirical investigations documented the presence of long-range dependence in four emerging eastern European markets, thereby rejecting evidence in favour of the efficient market hypothesis. Similar analysis using wavelet based methods to detect long memory in the returns of the Dow Jones Industrial average (DJI) were employed by Elder and Serletis (2007) where no evidence of long memory was detected, thereby supporting results from a vast number of studies that reject the presence of long memory in the developed markets of the U.S. However, the presence of long memory in the equity returns of some developed markets of Europe, the U.S., and Japan is documented in Ozdemir (2007). Furthermore, Ozun and Cifter (2007), also using a wavelet based estimator of long memory, found some evidence of long-range dependence in the returns of the Istanbul Stock Index, thereby rejecting the weak form efficiency of Istanbul share prices. Similarly, evidence of long memory in the equity markets of G7 countries is documented in Bilal and Nadhem (2009). On the other hand, Mariani et al (2010), using detrended fluctuation analysis and truncated Levy flight method, found evidence of long memory in several eastern European markets. However, among the countries that are part of the Organisation for Economic Co-operation and Development (OECD), long-memory, as investigated by Tolvi (2003), was only evidenced in the smaller equity markets of Denmark and Finland. Jefferis and Thupayagale (2008), using a long-memory variant of the GARCH model, investigated long memory behaviour of some select African equity markets and found evidence supporting the presence of long-memory in the developing markets of Botswana and Zimbabwe. The presence of long memory in the developing markets of Central and Eastern European countries (CEE) is documented in the studies of Jagric et al. (2006) and Kasman et al. (2009), where the presence of long memory in equity returns is specifically limited to the developing markets of Hungary, Czech, Slovenia and Croatia. Kristoufek and Vosvrda (2012) constructed a measure of efficiency by measuring the distance between an efficient case and a vector containing long memory and other measures of fractality. Long memory is evidenced in many developing and emerging markets whereas all developed markets show signs of efficiency, with the Japanese NIKKIEI leading all other developed markets in terms of efficiency. Cont (2005) attempted to identify economic intuition and mechanisms behind the existence of fractality and long memory in returns and returns volatility. The possible economic factors underlying the existence of long memory in volatility are, i) heterogeneous investment horizons of market agents, ii) evolutionary trading models that employ genetic algorithms, iii) market fluctuations arising out of investors' sudden switch between several trading strategies, and iv) the inactivity of investors, operating at certain time periods and market regimes, based on trading strategies or behavioral aspects. Vuorenmaa (2005) investigated the time-varying long memory of Nokia Oyj returns using the wavelet OLS method and found significantly strong long memory during the dot-com bubble period. Ozun and Ciftr (2007), demonstrating the superiority of wavelet OLS method as compared to the spectral long memory estimator of Geweke and Porter-Hudak (1983), found significant long memory in the returns of Istanbul stock exchange. Similarly, DiSario et al. (2008), on investigating the volatility structure of S&P 500 returns using the wavelet OLS method, found evidence of long-memory in the S&P 500 returns volatility. In the same vein as the aforementioned studies, Tan et al. (2012) while examining the fractal structure of emerging economies using wavelet OLS method demonstrated significant long memory in the returns of larger firms as compared to smaller firms. Likewise, Tan et al. (2014), using the wavelet estimator of Jensen (1999) and detrended fluctuation analysis, examined long memory behavior of equity returns and volatility of ten markets from both developing and developed economies. On the other hand, Power and Turvey (2010) investigated long memory structure of fourteen commodity futures using the Hurst estimator of Veitch and Abry (1999) and demonstrated long-range dependence in all commodities. Boubaker and Peguin-Feissolle (2013) proposed semiparametric wavelet base long memory estimators and demonstrated its superiority, with respect to several non-wavelet estimators, using simulation experiments. More recently, Pascoal and Monteiro (2014), while investigating the predictability of the Portuguese stock returns using wavelet estimators of long memory, fractal dimension and the Holder exponent, found no evidence of long memory in the PSI20 returns, thereby confirming the efficiency of the Portuguese equity market. This study, however, implements the wavelet based approaches of Abry and Veitch (1998) and Abry et al. (2003) to graphically examine the Hurst exponents of select equity returns using a log-scale wavelet plot.

## 1.3 Data

The empirical data consists of some select developed stock markets of France (CAC40), Germany (DAX), U.S. (S&P500), Great Britain (FTSE100), Switzerland (SMI), and the Eurozone (STOXX50). The data for emerging Asian economies constitute the stock markets of India (BSE30), Brazil (Brazil), Indonesia (JKSE), Pakistan (KSE100), China (SSE), and Malaysia (KLSE). The period of study ranges from 01-07-1997 to 20-01-2014 consisting of 4096 dyadic length observations making it suitable for various wavelet methods. Returns of all the stock indices are calculated by taking the first order logarithmic differences.

## 1.4 Methodology

Long memory process is associated with a slow power law decay of the autocorrelation function of a stationary process $x$. The covariance function $\gamma_x(k)$ of the long memory process $x$ takes the following form,

$$\gamma_x(k) \sim c_\gamma k^{-(2-2H)}, \quad k \to +\infty \tag{1.1}$$

where $c_\gamma$ is a positive constant and $H \in (0, 0.5)$. The Hurst parameter $H$ is used to measure the presence of long memory. The spectrum $\Gamma_x(\nu)$ of the long memory process $x$ is given by,

$$\Gamma_x(\nu) \sim c_f |\nu|^{1-2H}, \quad \nu \to 0 \tag{1.2}$$

where $\nu$ is the frequency, $c_f = \pi^{-1} c_\gamma \Lambda(2H-1)\sin(\pi - \pi H)$, and the Gamma function is given by $\Lambda$. This mathematical structure of long memory processes is the reason for its inclusion in a class of stochastic processes which have the $1/|\nu|^\alpha$ form. The property of long memory also finds some close association with the phenomenon of scale invariance, self-similarity and fractals. Hence, statistically self- similar processes like fractional Brownian motion (FBM) is closely related to long memory phenomenon.

Let $\gamma_0$ be an arbitrary reference frequency selected by the choice of $\psi_0$, the mother wavelet. The amount of energy in the signal during scaled time $2^j k$ and scaled frequency $2^{-j} \nu_0$ is measured by the squared absolute value of the detail wavelet coefficient $|d_x(j,k)|^2$. A wavelet based spectral estimator of Abry et al. (1993) is constructed by taking a time average of $|d_x(j,k)|^2$ at a given scale, and is given by,

$$\hat{\Gamma}_x(2^{-j}\nu_0) = \frac{1}{n_j} \sum_k |d_x(j,k)|^2 \tag{1.3}$$

where $n_j$ is the "number of wavelet coefficients" at level $j$, and $n_j = 2^{-j} n$, where $n$ is the data length. Therefore, $\hat{\Gamma}_x(\nu)$ captures the amount of energy that lies within a given bandwidth and around some frequency $\nu$. Hence, $\hat{\Gamma}_x(\nu)$ can be regarded as an estimator for the spectrum $\Gamma_x(\nu)$ of x. The wavelet based estimator of the Hurst exponent $\hat{H}$ is designed by performing a simple linear regression of $\log_2(\hat{\Gamma}_x(2^{-j}\nu_0))$ on $j$, i.e.,

$$\log_2(\hat{\Gamma}_x(2^{-j}\nu_0)) = \log_2\left(\frac{1}{n_j}\sum_k |d_x(j,k)|^2\right) = (2\hat{H}-1)j + \hat{c} \qquad (1.4)$$

where $\hat{c}$ estimates $\log_2(c_f \int |\nu|^{(1-2H)} |\Psi_0(\nu)|^2 d\nu)$, where $\Psi_0$ is the Fourier transform of the mother wavelet $\psi_0$. A weighted least square estimator is constructed by performing a WLS fit between the wavelet scales $j_1$ and $j_2$ which gives the estimator of the "*Hurst exponent*", H.

## 1.5 Empirical results

The presence of long memory in the volatility of select equity returns, as given by the absolute value of equity returns, is investigated by applying the wavelet based estimator of the Hurst exponent developed by Abry and Veitch (1998) and Abry et al. (2003).

This method enables one to graphically analyze long memory from the log-log plot of the wavelet regression which contains additional information about the fractal nature of equity returns. The logscale diagram is a plot of wavelet variance at each scale against the wavelet scale. Formally, the plot of the logarithm of $v_j = \frac{1}{n_j}\sum_{k=1}^{n_j}|d_X(j,k)|^2$ against the wavelet scale $j$ gives the *logscale* diagram. Here $n_j$ is the "number of wavelet coefficients" at scale $j$ and $d_X(j,k)$ is the wavelet details of the process $X(t)$. The visualization of the *logscale* diagram can help one detect regions of long-range dependence via the help of an *alignment region* in the graph. The range of wavelet scales where log ( $v_j$ ) falls on a straight line is known as the *alignment region* (Abry et al. 2003) and perfect alignment, mostly at higher scales, normally constitutes long memory. In the logscale plot, perfect alignment requires the red straight line to cross (or touch) the vertical lines depicting the confidence band in an upward sloping manner. If the alignment region includes the largest scales in the logscale plot, then the returns exhibit long-range dependence. Furthermore, the value of the self-similar parameter[3] $\alpha$ should lie in the interval (0, 1). Correspondingly, the value of the Hurst exponent $H$ should lie in the interval (0.5, 1) for the data to exhibit long memory. Figure 1.1 gives the logscale diagram[4] of the equity returns of select developed markets. It can be observed from the figure that straight line slopes downward and the corresponding Hurst exponents for all six developed markets of Europe and the U.S. lie within the interval (0, 0.5) indicating short-memory.

Figure 1.1 Logscale diagrams of equity returns from developed markets

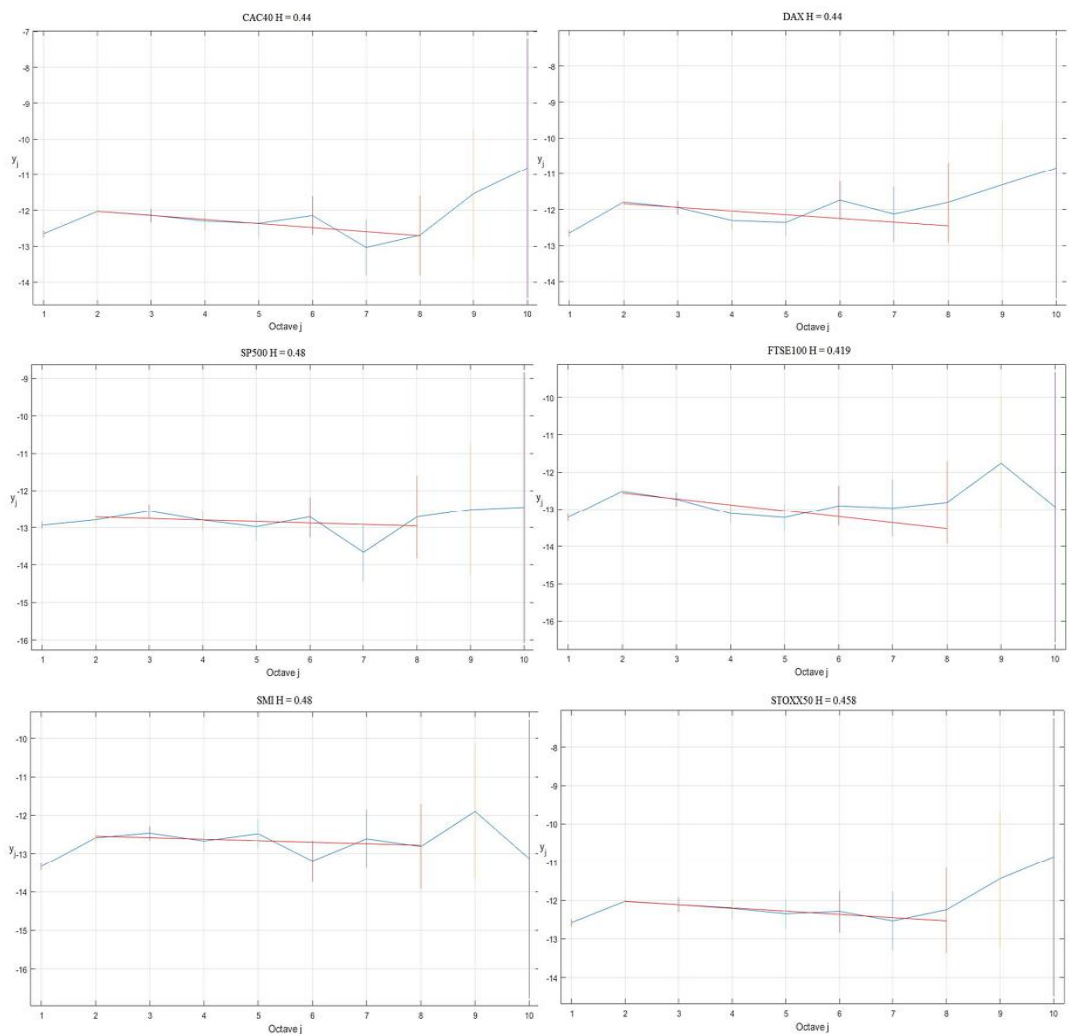

The absence of long memory in the returns of developed markets is in confirmation with results from a vast majority of literature that rejects long memory in developed financial markets. Figure 1.2 gives the logscale diagram of the equity returns of some select emerging markets. It can be noticed that the Hurst exponents of emerging markets' equity returns lie within the interval (0.5, 1).

---

[3] $\alpha$ is also known as the scaling exponent of self-similarity. The Hurst parameter $H$ and $\alpha$ are related by the expression: $H=(1+\alpha)/2$
[4] After repeated simulations, the optimal lower cut-off scale is taken to be 2 and the highest scale is taken to be 8.

Figure 1.2 Logscale diagrams of equity returns from emerging markets

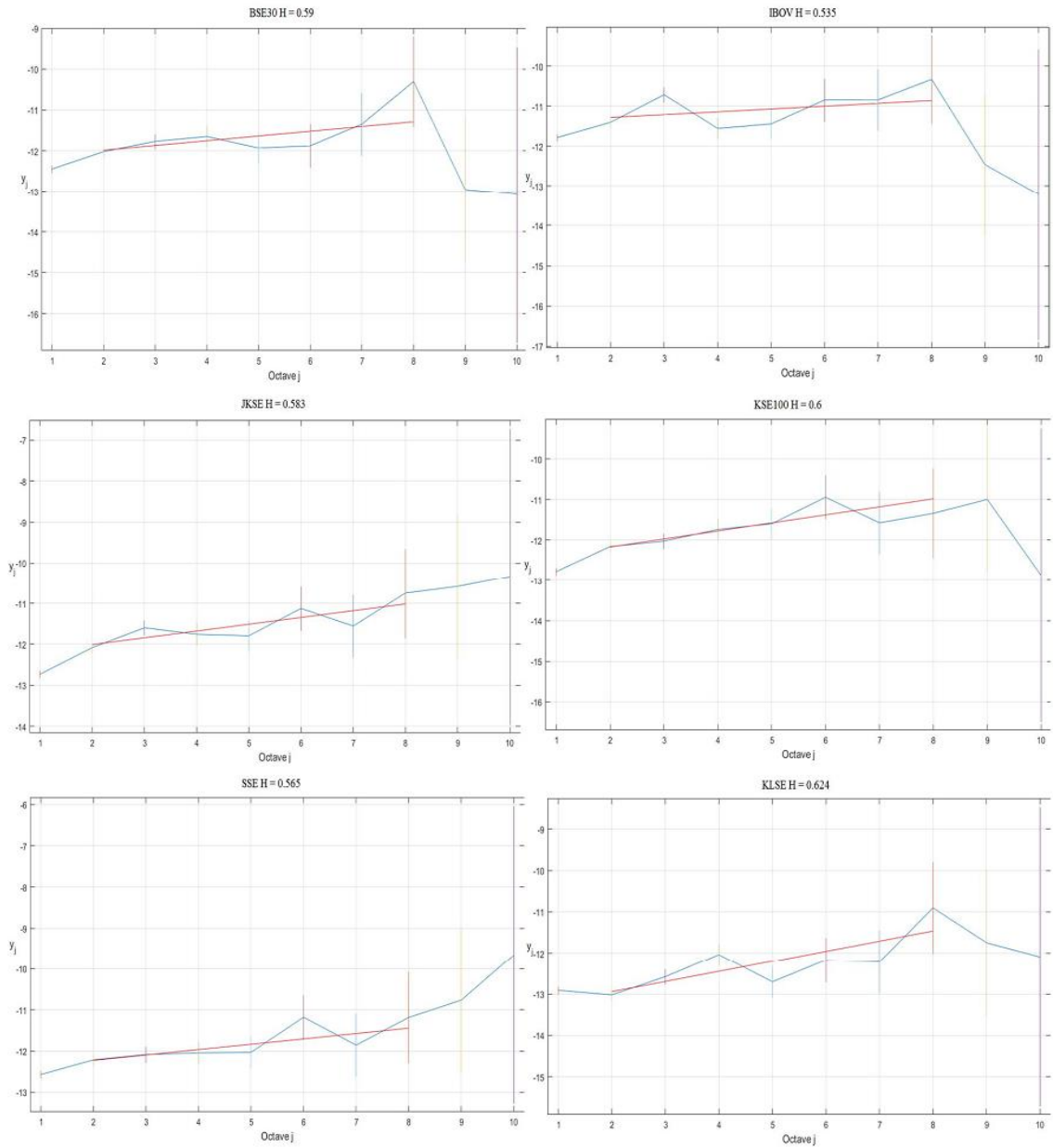

It can be observed from the above figure that the upward sloping alignment of the straight red line includes all higher scales, i.e. scales five up to eight, indicating the presence of "long-range dependence". However, among the six emerging markets, equity returns of India (BSE 30), Pakistan (KSE 30) and Malaysia (KLSE) exhibit relatively stronger long-memory.

## 1.6 Conclusion

This study investigated the phenomenon of long memory among select global equity returns using an improved method from the wavelet domain. Some evidence of long-memory in equity returns of emerging markets of Malaysia, Taiwan, Pakistan, China and Indonesia are unearthed. However, the application of improved fractal estimators of Abry et al. (2003), aided by the logscale diagram of wavelet based scaling estimates, detected significant long memory in the emerging markets of India, China and Indonesia. On the other hand, equity returns of developed markets from Europe and the U.S. did not exhibit long-range dependence, thus validating results from existing studies that reject long memory in developed markets. Moreover, markets from developed economies are said to be more efficient where efficiency is inversely related to the persistence of returns and prices. Therefore, the presence of long memory in equity returns rubbishes the notion of market efficiency. However, equity markets are in a constant stage of development which can influence efficiency and predictability.